\documentclass[12pt,preprint]{aastex}

\newcommand{\unit}[2]{\ensuremath{{#1} \, {#2}}}

\newcommand{\Mjup}{M_J}

\newcommand{\scinot}[2]{\ensuremath{#1 \times 10^{#2}}}

\newcommand{\paren}[1]{\left ( #1 \right )}

\newcommand{\parenfrac}[2]{\paren{\frac{#1}{#2}}}

\newcommand{\OmegaKep}[1]{\ensuremath{\Omega_{\textrm{kep}}^{#1}}}

\newcommand{\Mstar}[1]{\ensuremath{M_{*}^{#1}}}
\newcommand{\Mplanet}[1]{\ensuremath{M_{\textrm{p}}^{#1}}}

\newcommand{\OmegaPlanet}[1]{\ensuremath{\Omega_{\textrm{p}}^{#1}}}

\newcommand{\qdisc}[1]{\ensuremath{q_{\textrm{disc}}^{#1}}}

\begin{document}

\title{Giant Planet Migration in Viscous Power-Law Discs}

\author{Richard~.G.~Edgar}
\affil{Department of Physics and Astronomy, University of Rochester,
Rochester, NY 14627}
\email{rge21@pas.rochester.edu}


\begin{abstract}
Many extra-solar planets discovered over the past decade are gas giants in tight orbits around their host stars.
Due to the difficulties of forming these `hot Jupiters' in situ, they are generally assumed to have migrated to their present orbits through interactions with their nascent discs.
In this paper, we present a systematic study of giant planet migration in power law discs.
We find that the planetary migration rate is proportional to the disc surface density.
This is inconsistent with the assumption that the migration rate is simply the viscous drift speed of the disc.
However, this result can be obtained by balancing the angular momentum of the planet with the viscous torque in the disc.
We have verified that this result is not affected by adjusting the resolution of the grid, the smoothing length used, or the time at which the planet is released to migrate.
\end{abstract}



\section{Introduction}
\label{sec:intro}

The past decade has seen the discovery of over one hundred extra-solar planets.
Most of these planets \citep[see, e.g.][]{2005PThPS.158...24M} are gas giants of Jupiter mass or greater, in orbits close to their host stars.\footnote{For the most up to date information, refer to \url{http://exoplanet.eu/} and \url{http://exoplanets.org/}}
Such planets are often called `hot Jupiters.'
Although detection methods (primarily radial velocity measurements) are strongly biased towards detecting such planets, a sufficient number have been discovered to require detailed explanation.

The present orbits of the `hot Jupiters' lie so close to their host stars that in situ formation is almost impossible.
The problem is two-fold.
Firstly, the expected disc temperature is high, which suppresses gravitational capture of gas by a gas giant core.
Secondly, there simply isn't much material at such small orbital radii.
We are therefore led to the conclusion that the `hot Jupiters' formed further from the star, and migrated to their present orbits.
In this paper, we present a study of giant planet migration in power law discs.

A planet in a circumstellar disc exerts a torque on the disc material at resonances \citep{1978ApJ...222..850G,1979ApJ...233..857G,1980ApJ...241..425G}.
By Newton's Third Law, there is an opposite torque on the planet, which causes it to migrate.
Planets of mass comparable to Jupiter (the exact threshold depends on the disc conditions, see below) exert torques comparable to the viscous torque in the disc.
They can therefore open a gap in the disc.
If the gap is perfectly clean, then the planet will act as a `relay station' between the inner and outer discs.
We then expect the planet's orbital evolution to follow the viscous evolution of the disc -- a process known as Type II migration \citep{2000prpl.conf.1135W}.
We shall compare our computational results to the predictions of Type II theory.

We summarise the theory of Type II migration in Section~\ref{sec:typeii}.
We describe our numerical method in Section~\ref{sec:numerics}, and demonstrate the formation of a gap in Section~\ref{sec:gapdevel}.
Section~\ref{sec:results} presents our main results.
In Section~\ref{sec:discuss} we discuss the implications of our findings, before presenting our conclusions in Section~\ref{sec:conclude}.

\section{Type II Migration}
\label{sec:typeii}

In this section, we shall briefly review the theory of Type II migration.
For a more thorough analysis of the theory of planet--disc interactions, the reader should turn to one of the many published reviews \citep[e.g.][]{1993prpl.conf..749L,2000prpl.conf.1135W,2000prpl.conf.1111L}.

Planet--disc interactions are dominated by resonances.
A planet embedded in a circumstellar disc excites waves at its Lindblad resonances (LR).
These waves carry angular momentum, and hence exert a torque on the disc material.
The strength of the torque from the Lindblad resonances is proportional to $T_{\textrm{LR}} \propto \Sigma q^{2}$, where $\Sigma$ is the local disc surface density, and $q$ is the planet-star mass ratio.
These torques act to push material away from the planet.
At the same time, the disc gas is expected to have an intrinsic viscosity, $\nu$ (although the precise origin and exact behaviour of such a viscosity are still much debated), which leads to a viscous torque $T_{\nu} \propto \Sigma \nu$.
Since the Lindblad torques scale with $q^{2}$, we can expect that as the planet accretes material, the Lindblad torques will eventually dominate the viscous torque in the disc.
Balancing the two torques leads to the so-called gap opening criterion:
\begin{equation}
q > 40 \mathcal{R}^{-1}
\label{eq:gapopencriterion}
\end{equation}
where $\mathcal{R} = r^2 \Omega / \nu$ is the Reynolds number of the disc \citep{1999ApJ...514..344B,2000MNRAS.318...18N}.
There is a similar requirement that the planet's Hill sphere exceed the local scale height of the disc, namely that
\begin{equation}
q > 3 \parenfrac{h}{r}^{3}
\label{eq:tidalcondition}
\end{equation}
where $h$ is the disc scale height.
See \citet{1993prpl.conf..749L} for a further discussion.
The tidal condition of Equation~\ref{eq:tidalcondition} leads to the gap width being at least twice the local scale height.
If this condition is not satisfied, then the edge of the gap will be Rayleigh unstable.
Consideration of the torque condition leads to the expectation that the gap will lie between the $m=2$ Lindblad resonances of the planet (that is, between $r=0.6$ and $r=1.3$).
For Jovian mass planets in circumstellar discs, these gap sizes are similar.

Once the planet has opened a gap, it is assumed to isolate the inner and outer discs from each other.
However, each part of the disc will still be undergoing viscous evolution.
According to Type II migration theory, the planet will be locked to the viscous evolution of the disc \citep{2000prpl.conf.1135W}.
Setting the migration rate of the planet equal to the radial drift velocity of the gas in a thin accretion disc \citep[see e.g.][]{1981ARA&A..19..137P}, we find:
\begin{equation}
\dot{a} = -\frac{3 \nu}{2a}
\label{eq:TypeIIMigPredict}
\end{equation}
so long as the disc is sufficiently massive.
Note that Equation~\ref{eq:TypeIIMigPredict} is independent of the disc surface density.
If the disc mass is too low, then the torques (also proportional to $\Sigma$ above) will not be able to force the planet to migrate at this rate.
\citet{1995MNRAS.277..758S} and \citet{1999MNRAS.307...79I} have examined this limit.

\cite{2004ApJ...604..388I} suggest an alternative prescription for determining the Type II migration rate.
They balance the angular momentum change of the planet, $\frac{1}{2} \Mplanet{} \OmegaPlanet{} a \dot{a}$ with the maximum viscous couple in the disc $\dot{J} = \frac{3}{2} \Sigma \nu \OmegaKep{} r^2$.
Since we are using power law discs here, $\dot{J}$ will not have a maximum, so we use the nominal value at the planet's orbit.
This leads to the prediction
\begin{equation}
\dot{a} = - 3 \frac{\nu \Sigma a}{\Mplanet{}}
\label{eq:TypeIIAngMomBalance}
\end{equation}
Since this equation was obtained by balancing angular momentum, we might expect it to be valid for a full range of disc masses.

Surprisingly, giant planet migration does not seem to have been subjected to a systematic test (this is in sharp contrast to the theory of Type I migration, which applies to low mass planets).
Some curiosities in the behaviour of migrating giant planets have been seen \citep[e.g.][]{2004A&A...418..325S}, but these have not been explored in detail.
The work of \citet{2000MNRAS.318...18N} provides the most complete set of runs to date, but the physical parameters were not varied in a regular fashion.

In this paper, we shall present a series of numerical experiments, following giant planets migrating in a variety of accretion discs.
We will then compare our results to equations~\ref{eq:TypeIIMigPredict} and~\ref{eq:TypeIIAngMomBalance}.
Equation~\ref{eq:TypeIIMigPredict} makes particularly strong predictions about the expected migration rates - namely that the migration rate depends solely on the disc viscosity.
This is the first time such a test has been performed.

\section{Numerical Set up}
\label{sec:numerics}

We use the \textsc{Fargo} code of \citet{2000A&AS..141..165M,2000ASPC..219...75M} to perform our calculations.
\textsc{Fargo} is a simple \textsc{2d} polar mesh code dedicated to disc planet interactions. 
It is based upon a standard, \textsc{Zeus}-like \citep{1992ApJS...80..753S} hydrodynamic solver, but owes its name to the \textsc{Fargo} algorithm upon which the azimuthal advection is based.
This algorithm avoids the restrictive timestep typically imposed by the rapidly rotating inner regions of the disc, by permitting each annulus of cells to rotate at its local Keplerian velocity and stitching the results together again at the end of the timestep.
The use of the \textsc{Fargo} algorithm typically lifts the timestep by an order of magnitude, and therefore speeds up the calculation accordingly.
The mesh centre lies at the central star, so indirect terms coming from the planets and the disc are included in the potential calculation.
We make use of an non-reflecting inner boundary, to prevent reflected waves from interfering with the calculations.
The pitch angle of the wake is evaluated in the WKB approximation.
The inner ring of active cells is then copied to the ghost cells, with an azimuthal shift appropriate to the pitch angle.
Material which flows off the inner boundary is not added to the star (nor does the planet itself accrete).
At the outer boundary, mass was added, to compensate for the viscous evolution of the system.

We use units normalised such that $G=\Mstar{}+\Mplanet{}=1$, while the planet's initial orbital radius is set at $a=1$.
References to times in terms of `orbits' should be understood to mean ``orbital times at the planet's initial radius.''
The grid extends between $r=0.4$ and $r=2.5$.
Scaled to Jupiter's orbit, this grid roughly covers the area between the asteroid belt and Saturn's orbit.
We assume a constant aspect ratio disc, with $h/r = 0.05$.
We set the mass ratio $q=\Mplanet{}/\Mstar{}$ to be $10^{-3}$, approximately equal to the Jupiter-Sun value.
In our parameter space search, we varied the disc surface density profile and viscosity.
The surface density was initially a power law:
\begin{equation}
\Sigma (r) = \Sigma_0 \parenfrac{r}{r_0}^{-\delta}
\label{eq:discProfile}
\end{equation}
and we take $r_0 = 1$ (the planet's initial orbital radius).
Four values for $\delta$ are considered: 0, 0.5, 1 and 1.5.
These represent a reasonable range of alternatives, from a theoretically simple constant surface density disc to the canonical Minimum Mass Solar Nebula.
We set $\Sigma_0$ through
\begin{equation}
\qdisc{} = \frac{\Sigma_0 \pi a^2}{\Mstar{}}
\label{eq:qDiskDefine}
\end{equation}
which provides a quick \emph{estimate} of the disc's mass within the planet's orbit.
This estimate is accurate for $\delta=0$ discs, but is an underestimate for larger $\delta$ values.
We take four values for \qdisc{}: \scinot{5}{-4}, $10^{-3}$, \scinot{2}{-3}, and \scinot{3}{-3}.
The total disc mass lies between \scinot{1.9}{-3} (for $\qdisc{}=\scinot{5}{-4}$ and $\delta=1.5$) and $0.018$ (for $\qdisc{}=\scinot{3}{-3}$ and $\delta=0$) in units of the stellar mass.
By way of comparison, the Minimum Mass Solar Nebula (MMSN) requires at least \unit{5}{\Mjup} of gas in the vicinity of Jupiter's orbit.\footnote{We calculate this by comparing Jupiter's metal content to that of the Sun, and assuming that both condensed from the same gas cloud. Scaled to the Solar System, our grid roughly covers the region between the asteroid belt and Saturn}
Thus, our lower mass discs are somewhat sub-Minimum.

The viscosity is taken to be uniform, and has values $\nu = 10^{-4}$ and $10^{-5}$ in our units.
With a uniform viscosity, $\delta=0.5$ yields a disc with an initially stationary surface density profile \citep[cf equation~2.10 of][]{1981ARA&A..19..137P}.
These viscosities may be related to the $\alpha$ prescription for viscosity, $\nu = \alpha c_s h$ using
\begin{equation}
\nu = \alpha \parenfrac{h}{r}^2 \Omega r^2
\end{equation}
This implies that $\alpha$ varies with radius.
With our aspect ratio, a viscosity of $\nu = 10^{-5}$ gives $\alpha \approx \scinot{4}{-3}$ at the planet's initial orbital radius.
Note that for the highest viscosity, the gap opening criterion of Equation~\ref{eq:gapopencriterion} is not satisfied.
The tidal condition of Equation~\ref{eq:tidalcondition} is always satisfied in our numerical experiments.

The gravitational effect of the planet on the disc is smoothed at 0.6 of the disc thickness at the planet's orbital radius:
\begin{equation}
\phi = - \frac{G M}{\sqrt{r^2 + \epsilon^2}}
\end{equation}
where $\epsilon = 0.6 h$.
There are two motivations for this, the first being the desire to avoid having a singularity wandering around the grid.
The second is physical.
The \textsc{2d} approximation becomes poor close to the planet, where the vertical distribution of material becomes important.
The actual distance of material from the planet ceases to be the well approximated by the in-plane distance, which would lead to the gravitational effect being over-estimated.
Accordingly, we soften the potential over distances comparable to the disc scale height.
However, this softening length is still substantially smaller than the expected gap size and the planet's Hill sphere.
When calculating the torque the disc exerts on the planet, material from within the Hill sphere is subject to an exponential cut off, for similar reasons.
At the start of each run, the planet is introduced gradually (over about one orbit), and is not initially permitted to migrate.
This is done to minimise the effect of transients caused by the sudden appearance of a planet in a smooth disc.
We considered release times of approximately one orbit, and 100 \& 1000 orbits.
The computational grid is covered by 128 radial and 384 azimuthal cells (all uniformly spaced).

So far as possible, this setup mirrors that used in the comparison project presented by \citet{2006MNRAS.370..529D}.
In that comparison, the \textsc{Fargo} code was seen to give similar results to other codes used to study the disc--planet interaction problem.

\section{Development of the Gap}
\label{sec:gapdevel}

Since we introduce the planet into an initially unperturbed disc, there is a period of rapid evolution, as the planet clears a gap.
In this section, we shall discuss the development of this gap.

\begin{figure}
\begin{center}
\plotone{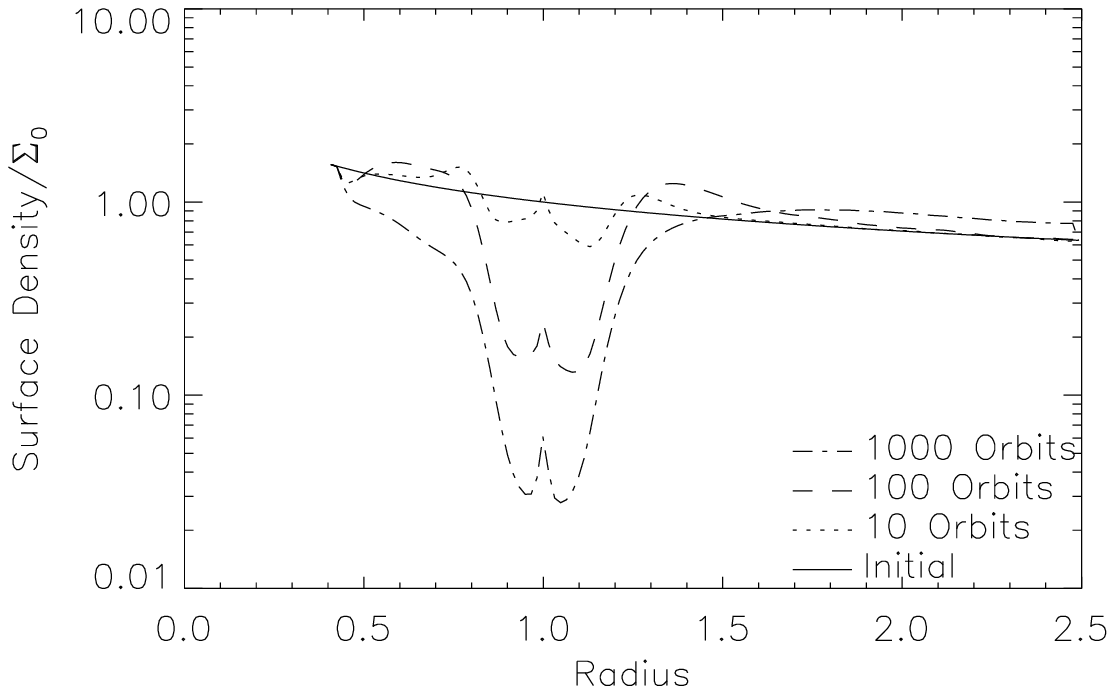}
\end{center}
\caption{Development of the gap for a Jupiter-mass planet on a fixed circular orbit, embedded in a $\nu=10^{-5}$ disc, with $\delta=0.5$ (cf equation~\ref{eq:discProfile}).
The azimuthally averaged surface density profile is show after 10, 100 and 1000 orbits.
Note that the $y$-axis is logarithmic}
\label{fig:GapEvolveMidVisc}
\end{figure}

In Figure~\ref{fig:GapEvolveMidVisc}, we trace the evolution of the gap in a $\nu=10^{-5}$ disc.
The initial surface density profile had $\delta=0.5$ (cf equation~\ref{eq:discProfile}).
There are no surprises in this plot, when compared with the many other numerical calculations of gap formation.
We see that the gap is mostly cleared in the first 100 orbits (note that the $y$-axis is logarithmic).
The gap lies roughly between the $m=2$ Lindblad resonances (located at $r=0.6$ and $r=1.3$).
For a $q=10^{-3}$ planet, this distance is also comparable to the co-rotation region.
This plot draws attention to the fact that the planet never completely clears the gap.
Even after 1000 orbits, the surface density in the gap is around 3\% of its initial value.
The gap edge is covered by roughly ten grid cells.

\begin{figure}
\begin{center}
\plotone{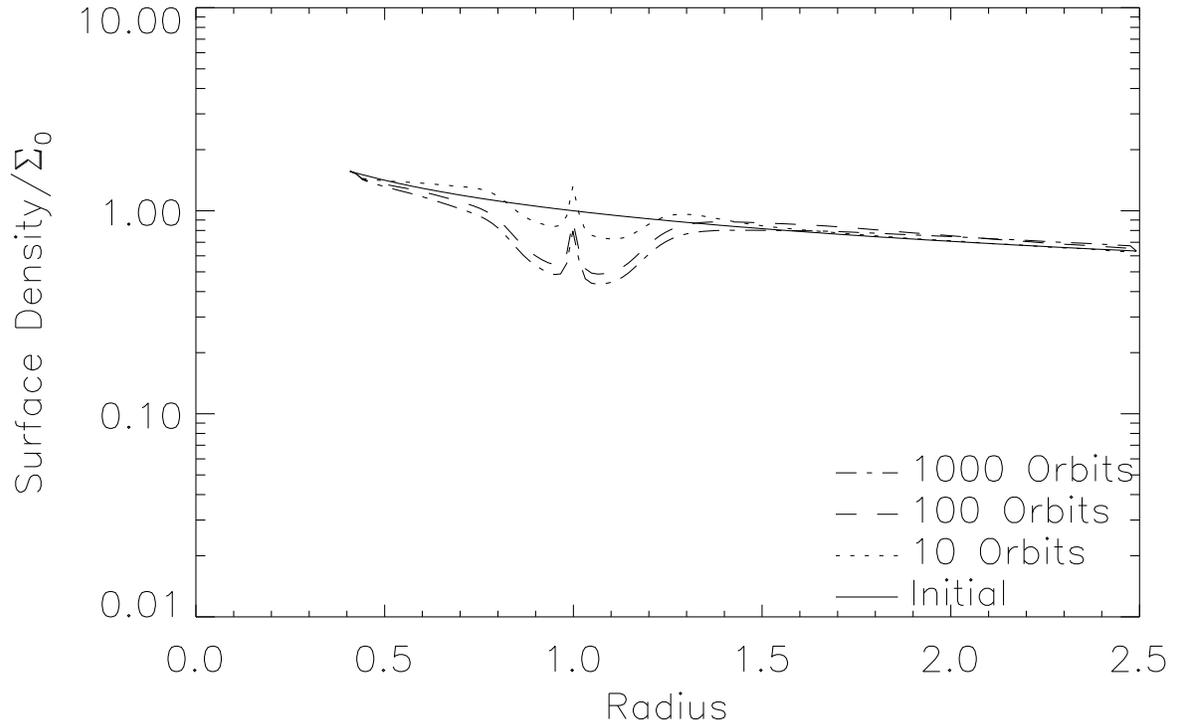}
\end{center}
\caption{Development of the gap for a Jupiter-mass planet on a fixed circular orbit, embedded in a $\nu=10^{-4}$ disc, with $\delta=0.5$ (cf equation~\ref{eq:discProfile}).
The azimuthally averaged surface density profile is show after 10, 100 and 1000 orbits.
For ease of comparison, the $y$-axis is identical to that in Figure~\ref{fig:GapEvolveMidVisc}}
\label{fig:GapEvolveHighVisc}
\end{figure}

If we increase the viscosity to $\nu=10^{-4}$, the gap becomes far less pronounced.
We show the development of the gap in this case in Figure~\ref{fig:GapEvolveHighVisc}.
Again, most of the depletion occurs during the first 100 orbits, but the total depletion is far less.
The density has only dropped to around 50\% of its initial value.
This is not unexpected - according to Equation~\ref{eq:gapopencriterion}, a Jupiter mass planet should not be able to open a gap in such a viscous disc.
Of course, the condition of Equation~\ref{eq:tidalcondition} is satisfied.

Figures~\ref{fig:GapEvolveMidVisc} and~\ref{fig:GapEvolveHighVisc} both show that most of the gap clearing occurs withing the first 100 orbits.
We shall therefore use this as our canonical release time below.
However, we shall show the effect of varying the release time as well.

Gaps in protoplanetary discs are known to vary smoothly with $q$ and $\nu$ (see, e.g. \texttt{astro-ph/0608020}), so what is taken to be a gap is somewhat arbitrary.
We shall continue with both viscosity values, but we must bear in mind that in the high viscosity case the gap is quite shallow.

\section{Results}
\label{sec:results}

We will now present the results of our numerical experiments of a Jupiter mass planet migrating in power law discs, grouped by viscosity.
Such planets are conventionally assumed to undergo Type II migration.
We have two predictions for the migration rate of giant planets in Equations~\ref{eq:TypeIIMigPredict} and~\ref{eq:TypeIIAngMomBalance}.
We shall compare our results to these predictions.


\subsection{High Viscosity}

The higher viscosity runs had $\nu = 10^{-4}$.
This viscosity means that for a Jupiter mass planet, the viscous gap opening criterion $q > 40 \mathcal{R}^{-1}$ of Equation~\ref{eq:gapopencriterion}  is not quite satisfied.
However, the tidal condition of Equation~\ref{eq:tidalcondition} is fulfilled.

\begin{figure*}
\plottwo{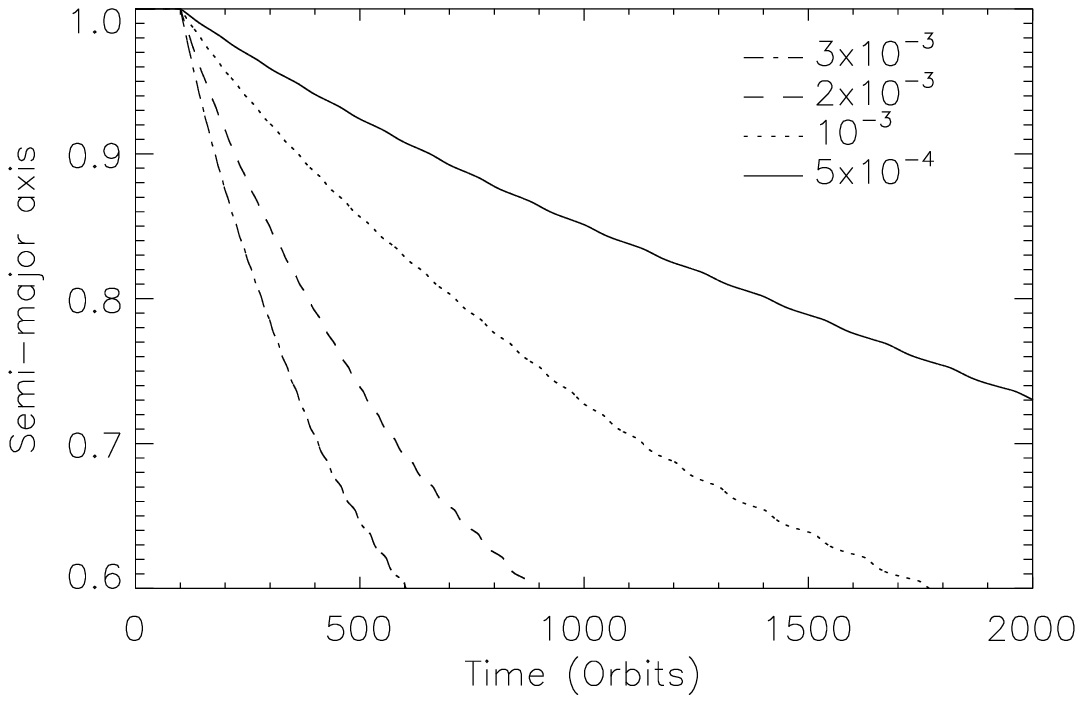}{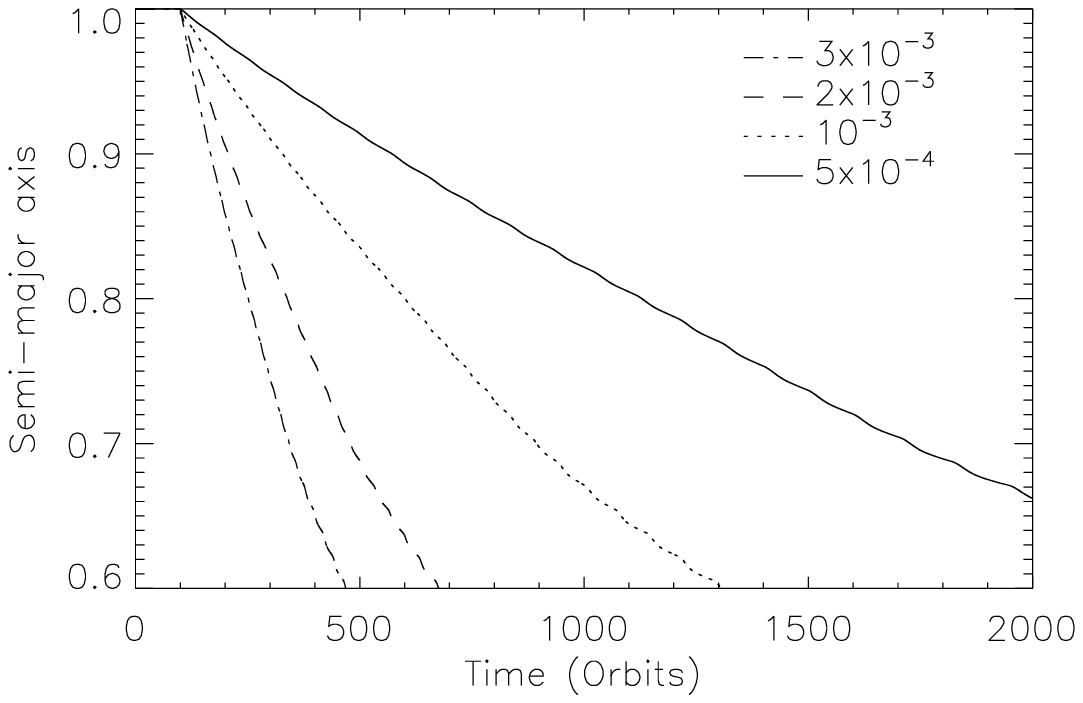} \\
\plottwo{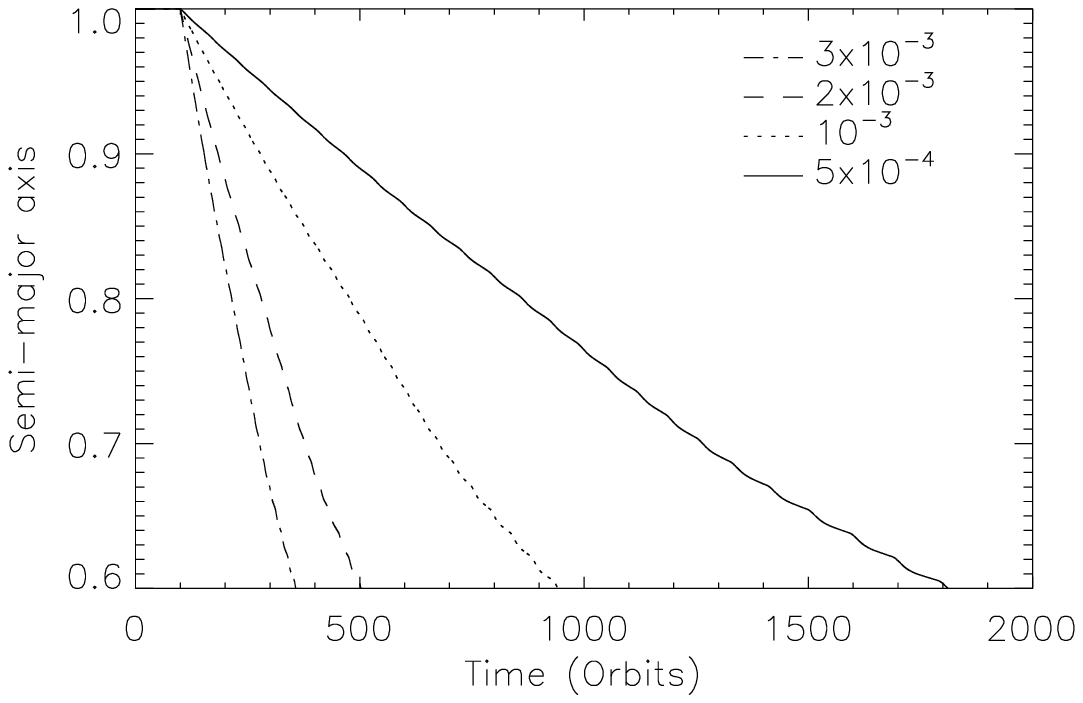}{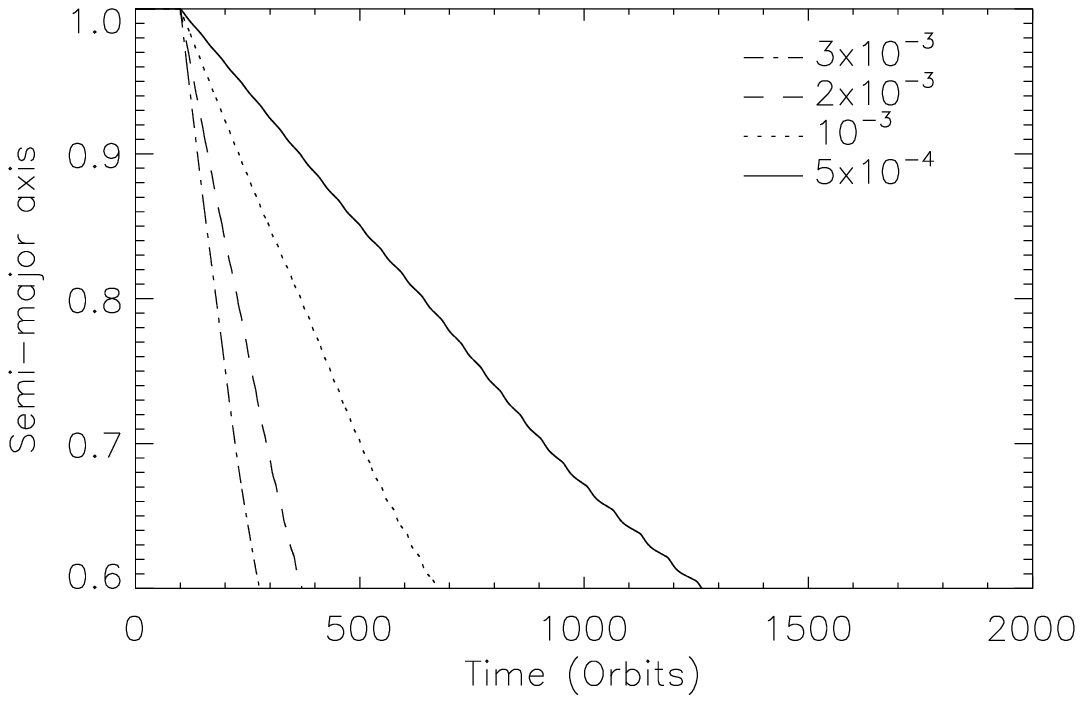}
\caption{Planetary migration in the $\nu=10^{-4}$ case for disc power laws of $\delta=0$ (top left), 0.5 (top right), 1.0 (bottom left) and 1.5 (bottom right). The planets were released to migrate after 100 orbits. The lines are marked by the value of \qdisc{} (see Equation~\ref{eq:qDiskDefine})}
\label{fig:HighViscMassCompare}
\end{figure*}

We shall discuss the results from the runs where the planet was released after 100 orbits.
In Appendix~\ref{sec:releasetime}, we demonstrate that the release time is not significant.
The orbital evolution of these planets is plotted in Figure~\ref{fig:HighViscMassCompare}.
We cut the $y$-axis at $0.6$, since at that point the $m=2$ ILR of the planet encounters the edge of the grid.
The migration rate of the planet therefore becomes unreliable.
We can see that the migration rate is a strong function of $\qdisc{}$ (or equivalently, $\Sigma_0$).
This is in direct contradiction to the prediction of Equation~\ref{eq:TypeIIMigPredict}.
Notice also how the migration rate varies with $a$.
Equation~\ref{eq:TypeIIMigPredict} predicts that $\dot{a} \propto a^{-1}$, which we do not see.
We see that the migration rate generally falls with $a$, which is consistent with equation~\ref{eq:TypeIIAngMomBalance}.
Figure~\ref{fig:HighViscMassCompare} shows that the reduction of $\dot{a}$ with $a$ falls as $\delta$ increases (that is, there is a pronounced curve in the migrations for $\delta=0$, whereas those for $\delta=1.5$ are almost straight lines).
This is broadly consistent with equation~\ref{eq:TypeIIAngMomBalance}, were the migration rate is proportional to $\dot{a} \propto a \Sigma \equiv a^{1-\delta}$.
Complicating this is the viscous evolution of the disc itself, which is probably the reason why $\dot{a}$ is not strictly proportional to $a^{1-\delta}$ (which would predict accelerating migration for the $\delta=1.5$ case).

We have seen that the migration rate of a Jupiter mass planet in these discs is strongly affected by the disc surface density.
However, since the gap is not particularly deep, whether Type II behaviour should be expected is debatable.


\subsection{Medium Viscosity}

Here, we examine the results from runs with $\nu = 10^{-5}$.
Again, we shall examine the case where the planet was released after 100 orbits first.

In Figure~\ref{fig:MidViscMassCompare}, we show the orbital migration for planets embedded in a variety of discs.
We see that planets embedded in higher surface density discs (parameterised by \qdisc{} -- cf Equation~\ref{eq:qDiskDefine}) consistently undergo faster migration.
The migration rate is roughly proportional to the disc surface density.
Note also the variation of $\dot{a}$ with $a$.
It is similar to that seen for figure~\ref{fig:HighViscMassCompare} above.
Eccentricities again remained low.


\subsection{Summary of results}

In this section, we have presented a series of giant planet migration runs.
Such planets have generally been thought to undergo Type II migration.
One formulation of the theory predicts that the migration rate depends solely on the disc viscosity (Equation~\ref{eq:TypeIIMigPredict}).
However, we have found that the migration rates vary systematically with disc surface density.
Higher disc surface densities give faster migration, which is consistent with Equation~\ref{eq:TypeIIAngMomBalance}.
There is a weaker variation with disc viscosity, which is inconsistent with both predictions.

In Appendix~\ref{sec:releasetime}, we show that our conclusions are not affected by varying the time at which the planet was released to migrate.
We demonstrate that the grid resolution does not affect our results in Appendix~\ref{sec:resolutionTest}.

\section{Discussion}
\label{sec:discuss}

In Section~\ref{sec:results}, we presented a series of runs designed to study how giant planets migrate.
This migration of such planets is thought to be controlled by the viscous torque within the disc.
Two different rates have been suggested.
In the first (Equation~\ref{eq:TypeIIMigPredict}), the planet is locked to the viscous evolution of the disc, and the migration rate depends solely on the disc viscosity.
The second (Equation~\ref{eq:TypeIIAngMomBalance}) computes an angular momentum balance between the planet and disc.
In this theory, the migration rate also depends on the disc surface density and planet mass.

We have seen that the migration rate we obtain varies strongly with disc surface density, indicating that Equation~\ref{eq:TypeIIMigPredict} in not appropriate.
Although Equation~\ref{eq:TypeIIAngMomBalance} is more promising, we do not recover the same variation with viscosity.
The higher viscosity runs underwent more rapid migration, but the difference in migration rates was not an order of magnitude.
However, this result is not as robust as the variation with surface density, since the high viscosity runs did not satify both gap opening criteria.
Although Figure~\ref{fig:GapEvolveHighVisc} shows that the high viscosity runs are definitely in the non-linear regime, the gap itself was not especially clean.

We have shown (Appendix~\ref{sec:releasetime}) that our results are not simply `turn-on' transients.
The migration rates are not significantly affected by the time at which the planet is released from a fixed orbit.
Our resolution tests (Appendix~\ref{sec:resolutionTest}) demonstrate that our results are not significantly affected by a doubling of the grid resolution.

Our neglect of material within the Hill sphere when calculating the torque is a point of concern.
\citet{2003ApJ...586..540D} noted that most of the torque in their calculations came from within the Hill sphere.\footnote{However, they did not perform a parameter space search like we have done here}
However, the theory of Type II migration takes no account of this material either.
It is a simple \textsc{2d} theory, which assumes that the planet is merely acting as a `relay station' for the disc's viscous torques.
If the flow structure within the Hill sphere is of critical importance, then we should not expect giant planet migration to be as simple as Section~\ref{sec:typeii} suggests.

The accretion behaviour of the planet could also affect migration.
This is directly linked to the previous point about flow within the Hill sphere.
\citet{1999MNRAS.303..696K} showed that even in the presence of a gap, a planet could continue to accrete material from the disc.\footnote{Note that this finding in itself implicitly contradicted the usual assumption that the planet isolates the inner and outer discs}
Similar results were reported by \cite{1999ApJ...526.1001L} and \citet{2001ApJ...547..457K}.
In this paper, we did not allow the planet to accrete, and this caused material to build up around the planet.
Since we attentuated the torque from within the Hill sphere, this would not have affected our results directly.
However, if accretion were allowed, then the planet could gain an appreciable amount of mass.
This would both alter the gap structure, and make it more difficult for the disc to move the planet (due to the planet's increased inertia).
Related to this issue is the recent finding of \citet{2006ApJ...641..526L} that the accretion rate through the gap could be over 10\% of the viscous accretion rate in the main disc, despite the drop in gas density.

Finally, there is the matter of viscosity.
In our numerical experiments, we used a physical viscosity in the Navier-Stokes equations.
In reality, the `viscosity' in protoplanetary discs probably originates from MHD turbulence, and calculations have shown \citep{2003ApJ...589..543W} that the gap structure obtained in an MHD calculation differs from that in a purely hydrodynamic one.
In particular, the gaps tend to be wider and shallower.
If material in the corotation region is important to determining the migration rate, then this alteration in gap structure will cause further changes to the migration rates.
\citet{2005A&A...443.1067N} has already demonstrated that a magnetic turbulence strongly affects the migration of a low mass planet.
Although our computations do not include MHD turbulence, the theory of Type II migration neglects it too, so this cannot be the reason for the differences we have observed.

When might we expect migration to proceed according to equation~\ref{eq:TypeIIMigPredict}?
We believe this might be possible for a planet of moderate mass, in a cold, very low viscosity disc, which is more massive than the planet.
Our reasoning is as follows:
equation~\ref{eq:TypeIIMigPredict} is based on the assumption that the planet completely isolates the inner and outer discs.
This is easiest to achieve in a very low viscosity disc (cf equation~\ref{eq:gapopencriterion}), which is also cold (cf equation~\ref{eq:tidalcondition}).
We also require the disc to be more massive than the planet, to ensure sufficient angular momentum reserves are available.
The gap will lie roughly between the $m=2$ Lindblad resonances, and we would want these particular resonances to be responsible for most of the gap clearing (i.e. the $m=2$ resonances themselves must dominate the disc's viscous torque).
This is because we would need disc material to be kept well away from the corotation region of the planet.
\citet{2003ApJ...588..494M} showed that corotation torques can give rise to extremely rapid migration - known as `runaway' or Type III migration.
A planet less massive than Jupiter will have its corotation region inside its $m=2$ Lindblad resonances.
However, \citeauthor{2003ApJ...588..494M} found that such planets tended to undergo runaway migration.
This reinforces the need for the disc itself to have a very low viscosity, so that the gap is as clean as possible.


\subsection{Origin of the Torque}
\label{sec:TorqueOrigin}

We shall now discuss the radial origin of the torque.

In figure~\ref{fig:TorqueProfileFlatProfileEvolve}, we show the torque profiles, $T_{z}(r)$ acting on the planet after 100, 500 and 1000 orbits.
The disc viscosity was $\nu=10^{-5}$, and the surface density profile was initially flat ($\delta=0$).
The planet was held on a fixed orbit for the entire calculation.
In computing the torques, the same exponential cut off used with \textsc{Fargo} was applied.
We see that most of the torque is generated within the range $0.8 < r < 1.2$, and that the torques from the inner and outer discs have opposite signs.
At later times, the torque peaks on either side of the planet lessen.
This is due to the gap emptying further.
The outer peak (which is pushing the planet inwards) also broadens, while the inner peak does not.
This ultimately ensures that the planet migrates inwards.

\begin{figure}
\plotone{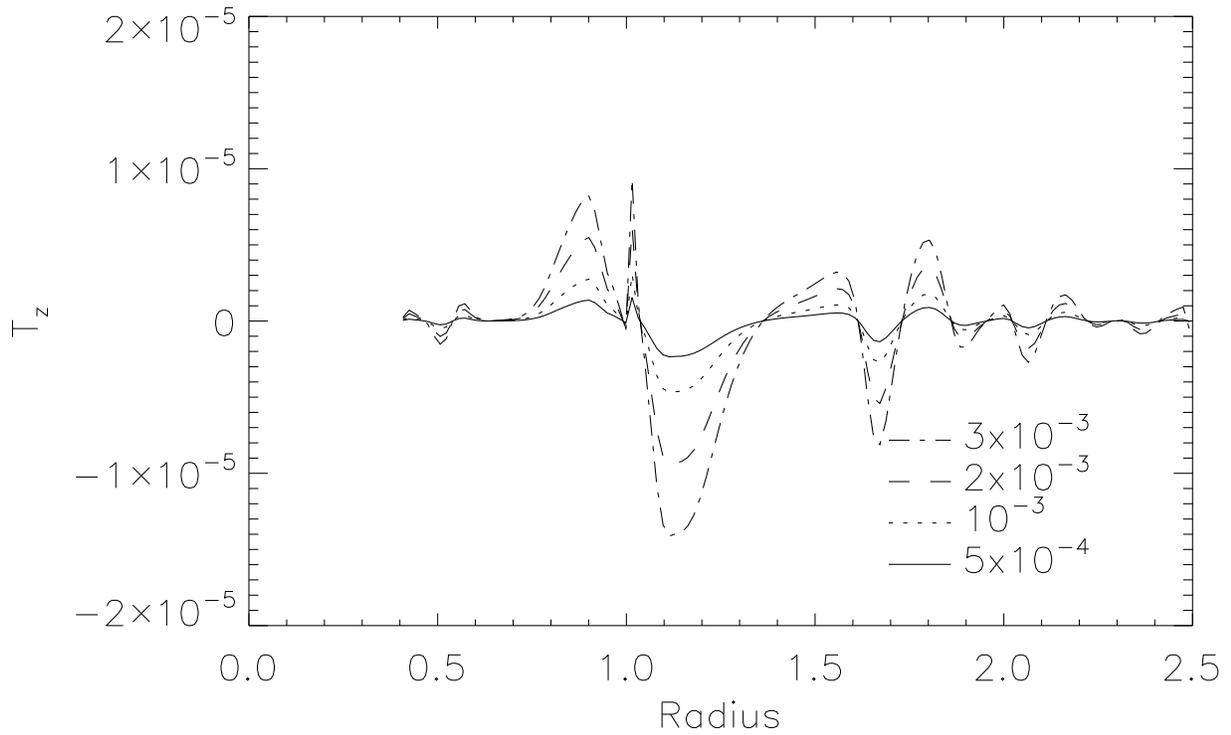}
\caption{Radial torque profiles for a planet in $\nu=10^{-5}$ discs of differing surface density after 1000 orbits.
The disc had a $\delta=0$ initial surface density profile, and each line is marked with its $\qdisc{}$ value.
The planet's orbit was fixed}
\label{fig:TorqueProfileFlatProfileMassVary}
\end{figure}

Figure~\ref{fig:TorqueProfileFlatProfileMassVary} shows how the torque felt by the planet scales with $\qdisc{} \propto \Sigma_0$.
These curves are plotted for a $\delta=0$, $\nu=10^{-5}$ disc after 1000 orbits.
As we might expect from section~\ref{sec:results}, we see that the strength of the torques is directly proportional to the value of $\qdisc{}$.
This leads to the migration rate varying strongly with the surface density of the disc.

\begin{figure}
\plotone{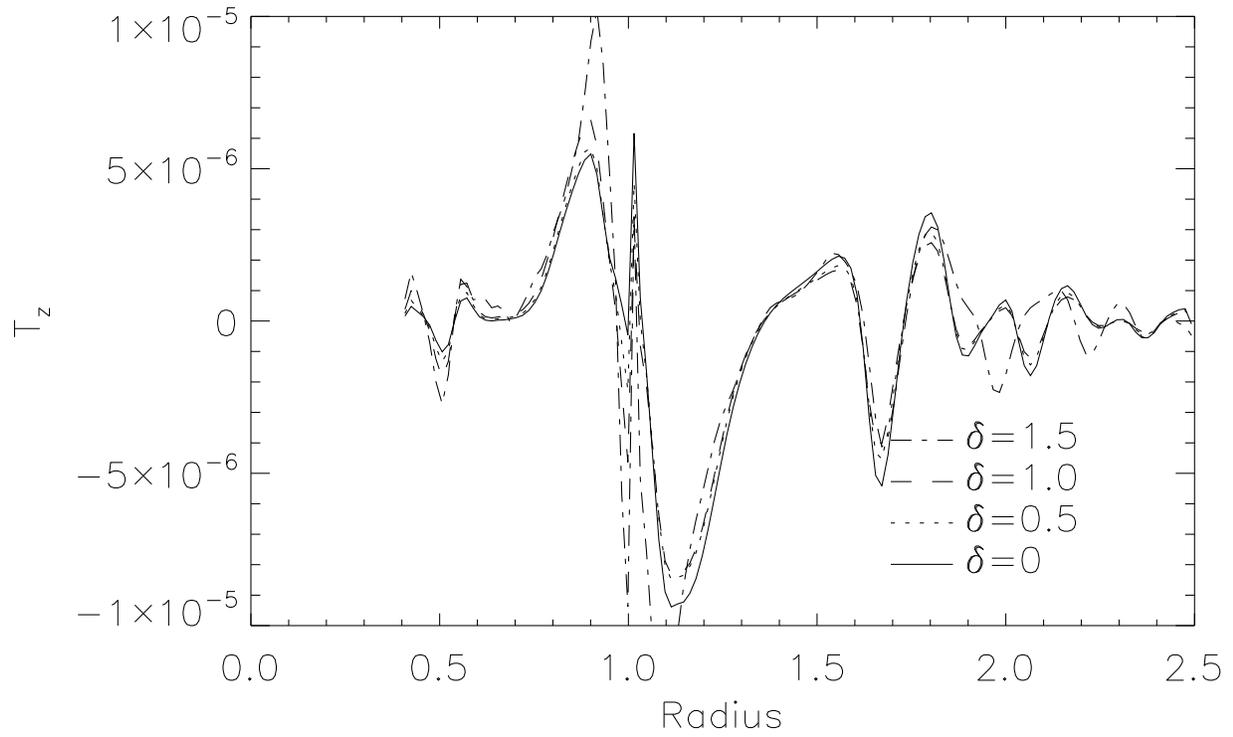}
\caption{Radial torque profiles for a planet in $\nu=10^{-5}$ discs of differing $\delta$, plotted after 1000 orbits.
All discs had the same $\qdisc{}$ value, and the planet was held on a fixed orbit}
\label{fig:TorqueProfileVaryDelta}
\end{figure}

Finally, in figure~\ref{fig:TorqueProfileVaryDelta}, we show the effect of varying the initial disc power law, $\delta$, on the torque profiles.
Again, the torque profiles are for a $\nu=10^{-5}$ disc, and are plotted after 1000 orbits of the (fixed) planet.
We see that the torques are very similar, regardless of the initial $\delta$ value, indicating that the perturbations induced by the planet are not dependent on the background structure of the disc.
Again, this is expected from section~\ref{sec:results}.

Figures~\ref{fig:TorqueProfileFlatProfileEvolve}, \ref{fig:TorqueProfileFlatProfileMassVary} and \ref{fig:TorqueProfileVaryDelta} all show that the torque generation peaks at radii of $r=0.9$ and $r=1.1$ (roughly 1.5 Hill radii from the planet).
Comparing to figure~\ref{fig:GapEvolveMidVisc}, we see that these locations lie deep within the gap, which is interesting for a number of reasons.
The peaks are close to the cutoff radius generally applied to obtain the numerical factor in equation~\ref{eq:gapopencriterion}, namely the planet's Hill sphere.
They are also well within the corotation region, raising the possibility that corotation torques are affecting the orbital evolution of the planet.
Unfortunately, at this resolution, the Hill radius is only covered by four or five grid cells, and the torque peak only lies seven grid cells from the planet itself.
The smoothing lengths are also comparable to these distances.

Our resolution tests (Appendix~\ref{sec:resolutionTest}) show that our resolution is adequate for the smoothing lengths used.
However, with so much torque being generated close to the planet, it is likely that the smoothing is significantly affecting the torque.
Reduction of the smoothing lengths is obviously desirable, but unfortunately not possible in a two dimensional calculation.
As noted in section~\ref{sec:numerics}, we must smooth the planet's gravity at about the local scale height of the disc in order to make a \textsc{2d} calculation valid.
In Appendix~\ref{sec:exclusionTest}, we show the effect of reducing the exclusion radius for the calculation of the migration torque.
The effect appears to be minimal, but these results should be treated with some caution, due to the issue noted above.
With the structure of the flow close to the planet so obviously critical to determining the migration rate, a full \textsc{3d} calculation would be required to determine a robust migration rate.
However, we would be most surprised if such calculations undermined our main conclusion that migration rates of massive planets are proportional to the disc mass.

\section{Conclusion}
\label{sec:conclude}

In this paper, we have performed a systematic test of giant (Jupiter mass) planet migration.
The migration rates we obtained varied strongly with the initial disc surface density, and less strongly with the disc viscosity.
We have shown that the simplest theory of Type II migration, where the planet is locked to the viscous evolution of the disc (Equation~\ref{eq:TypeIIMigPredict}), is incorrect.
An alternative formulation, based on an angular momentum balance (Equation~\ref{eq:TypeIIAngMomBalance}), looks more promising.
However, we have not tested this second theory fully.
We verified that our results were not simply `turn-on' transients, or purely the effect of low resolution.
Neither doubling the grid resolution, nor allowing the planet to clear its gap for 1000 orbits affected our central finding.

Separate confirmation of our results, using a different code would be highly desirable.
Although we have no reason to believe that \textsc{Fargo} is misbehaving, the work of \citet{2006MNRAS.370..529D} underlines how codes can give varying results, even for the `same' physical scenario (it is for this reason that `simulations' should properly be referred to as `numerical experiments').
The issues of accretion and gravitational softening (both of the planet's effect on the disc, and the torque exerted on the planet) also merit closer consideration.
Indeed, if the gap shape and flow through the gap are critical for migration, one is led to wonder if \textsc{2d} calculations are sufficient.
Two dimensional calculations have generally been thought adequate for Jovian mass planets because the gap would keep material away from the planet (where the \textsc{3d} nature of the flow will become evident).
If the flow within the Hill sphere is important, then \textsc{2d} calculations cease to be convincing.


\appendix

\section{Effect of Varying Release Time}
\label{sec:releasetime}

In this Appendix, we shall demonstrate that varying the time at which the planet is released does not affect our central conclusion.
We start by showing that changing the release time only affects the migration rates slightly.
We then show that, since the effect is consistent for all discs used, the conclusion that the planet migration rate varies with disc surface density is robust.

\begin{figure}
\begin{center}
\plotone{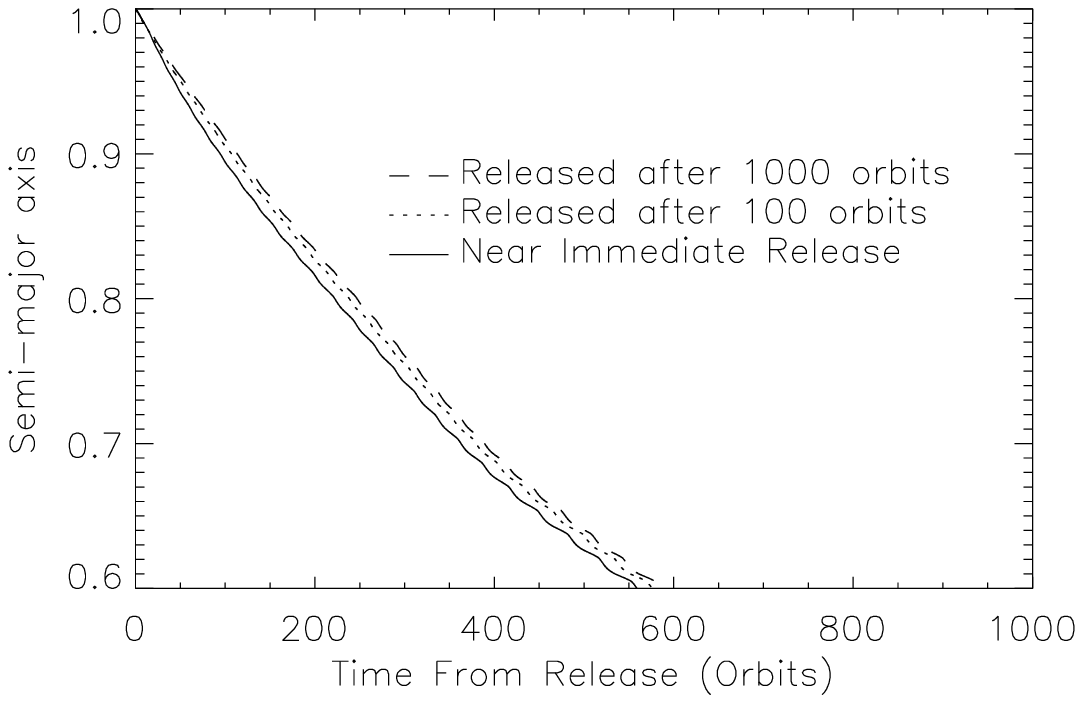}
\end{center}
\caption{The effect of release time on the $\nu=10^{-4}$ case.
The orbital evolution of a planet in a sample disc is shown for three different release times}
\label{fig:HighViscReleaseCompare}
\end{figure}

In Figure~\ref{fig:HighViscReleaseCompare}, we show the effect of varying the release time on a planet in a high viscosity ($\nu=10^{-4}$) disc with $\qdisc{}=0.002$ and $\delta=0.5$.
The effect is fairly minimal, and this plot is typical.
The explanation for this lies in Figure~\ref{fig:GapEvolveHighVisc}, which shows that only a minimal gap is formed.
Indeed, one can debate whether the surface density depression is a gap, since the usual criterion (Equation~\ref{eq:gapopencriterion}) is not satisfied.

\begin{figure}
\begin{center}
\plotone{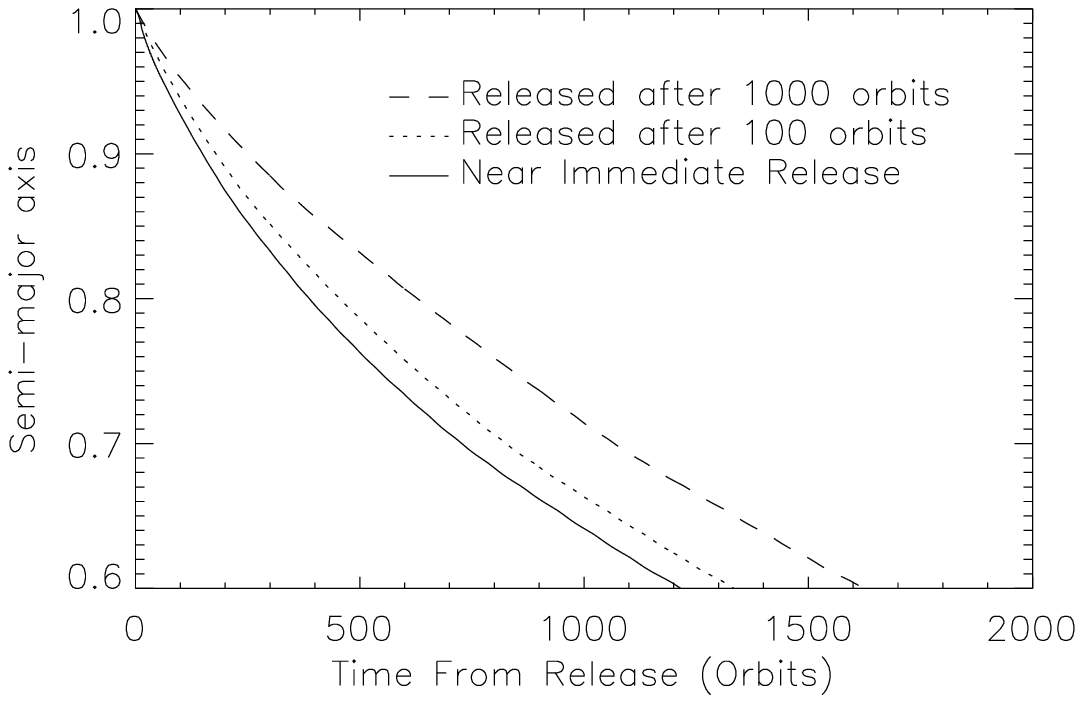}
\end{center}
\caption{The effect of release time on the $\nu=10^{-5}$ case.
The orbital evolution of a planet in a sample disc is shown for three different release times}
\label{fig:MidViscReleaseCompare}
\end{figure}

Figure~\ref{fig:MidViscReleaseCompare} shows the effect of release time on the orbital migration of a planet in the $\nu=10^{-5}$ case.
The particular disc used in this comparison had $\qdisc{} = 0.002$ and $\delta=0$, but the behaviour was generically the same for all cases.
We see that the time at which the planet is released does have an effect on the migration rate.
However, the effect is not especially dramatic.

We demonstrate that the release time does not affect our central conclusions in Figure~\ref{fig:MidViscMassCompareLongRelease}.
This duplicates Figure~\ref{fig:MidViscMassCompare}, but with the release time increased to 1000 orbits (note that the $x$ axis has the zero point shifted, to improve the use of space).
Although the exact migration rates undoubtedly change, the main conclusion that these planets are not undergoing Type II migration is unaffected.
The migration rates continue to be affected by the disc surface density.


\section{The Effect of Resolution}
\label{sec:resolutionTest}

In this appendix, we shall study the effect of increasing the grid resolution on our results.

We re-ran four of our numerical experiments, but with the grid resolution doubled.
We picked the set of four runs with $\nu=10^{-5}$, and $\delta=0.5$.
In Figure~\ref{fig:MidViscResolutionCompare} we plot the results, compared to the top right panel of Figure~\ref{fig:MidViscMassCompare}.
In this plot, we can see that doubling the grid resolution has a minimal effect on the migration rates obtained.

From this, we see that our conclusions are not simply an artifact of low resolution.
Of course, the smoothing we have used could be hiding some effects, but we would not expect decreasing the smoothing length to change the planet migration to Type II behaviour.
The smoothing length, at $0.6h$ is already rather smaller than the gap width, so shrinking it further is unlikely to enable the planet to make the gap cleaner.
Furthermore, this smoothing length is already as small as we can realistically make it.
The flow close to the planet will really have a \textsc{3d} structure, not resolved in these calculations.
By forcing all material to lie in the disc plane (as required by a \textsc{2d} calculation), we effectively make it closer to the planet -- significantly so within the gap.
By softening the potential over a distance comparable to the scale height, we approximate the true \textsc{3d} strength of the planet's gravity.


\section{The Effect of Hill Sphere Exclusion}
\label{sec:exclusionTest}

By excluding material within the planet's Hill sphere when computing the migration torque, we potentially reduced the migration torque substantially.
Although this has a sound physical motivation (cf section~\ref{sec:numerics}), it does potentially affect the migration rate of the planet.

\begin{figure*}
\plottwo{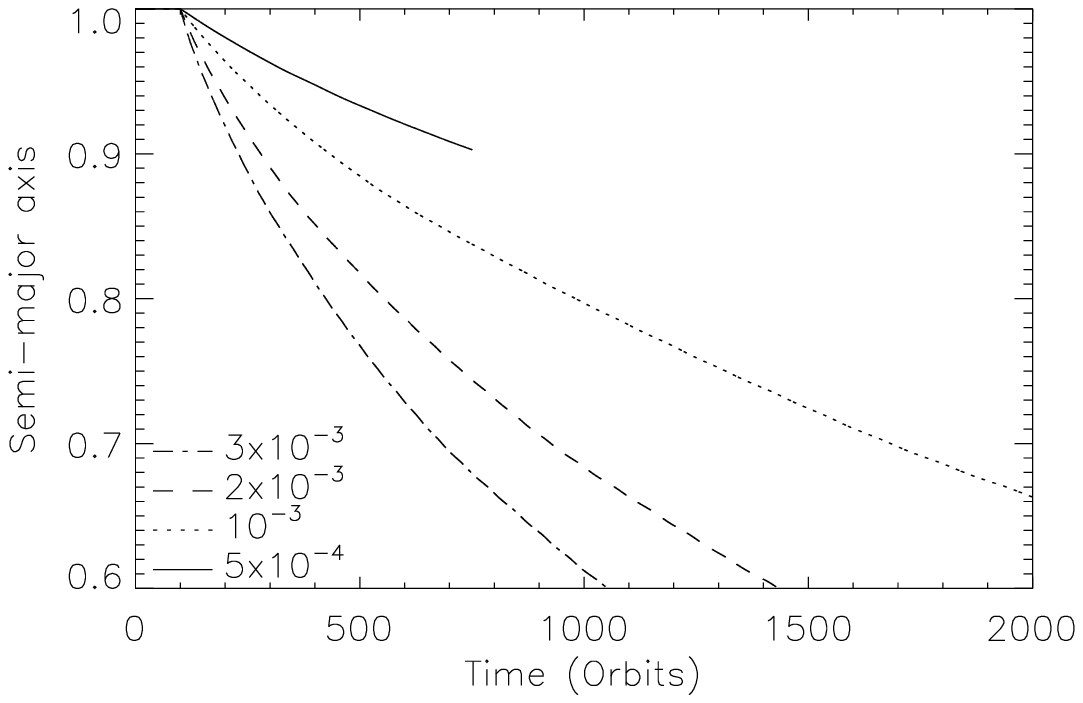}{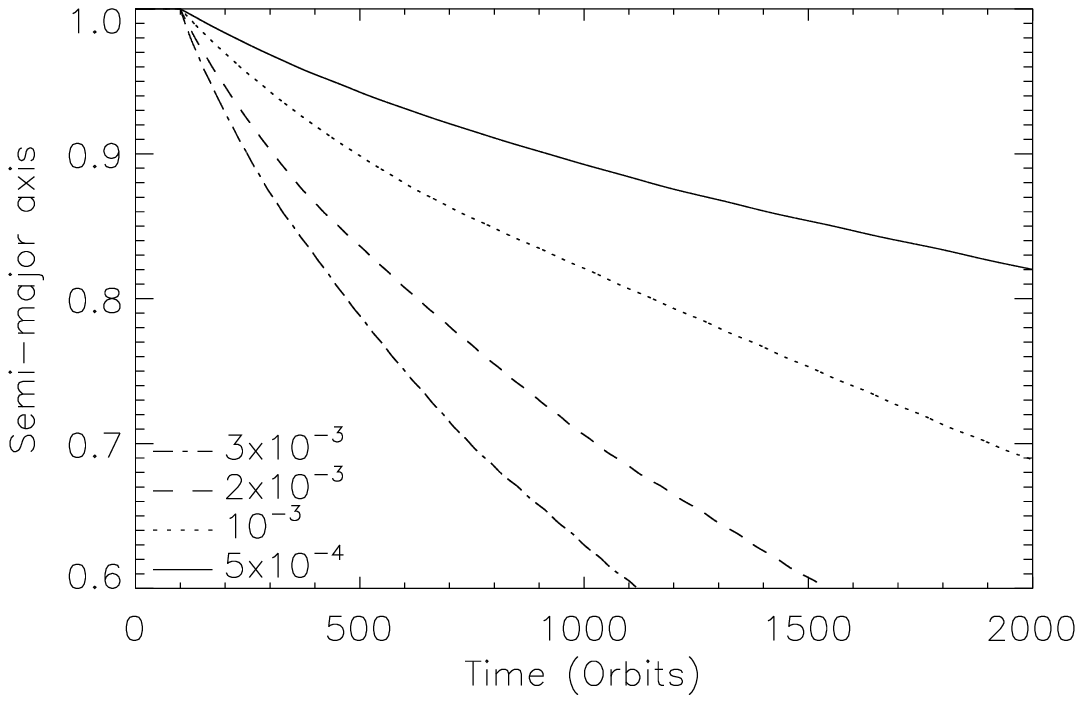} \\
\plottwo{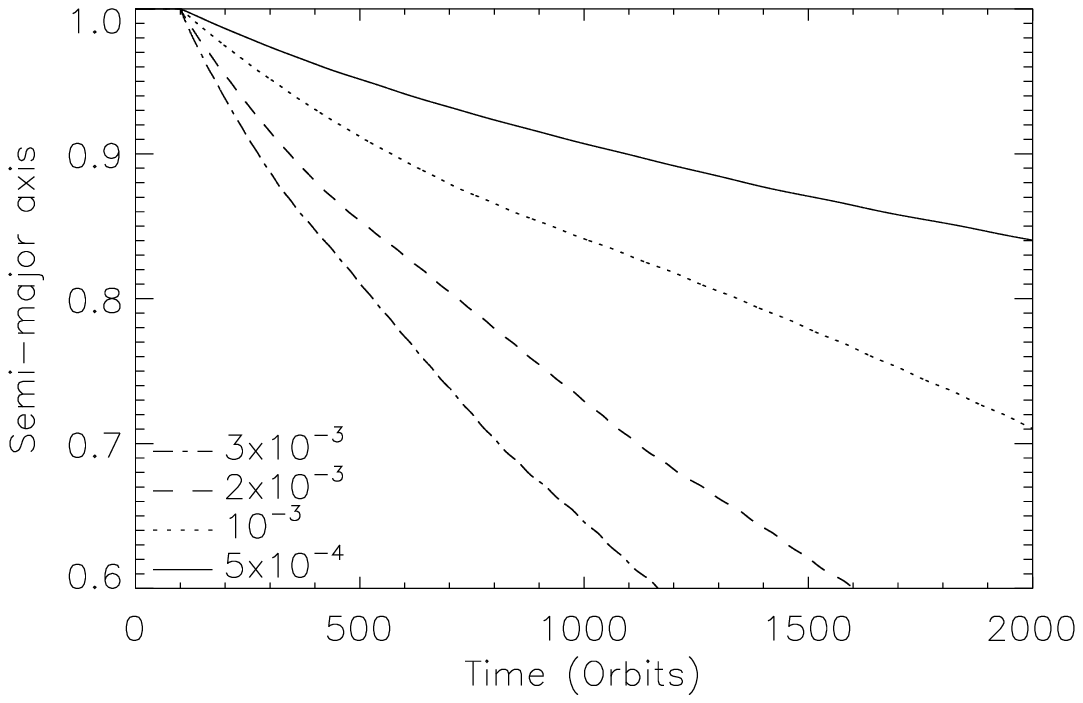}{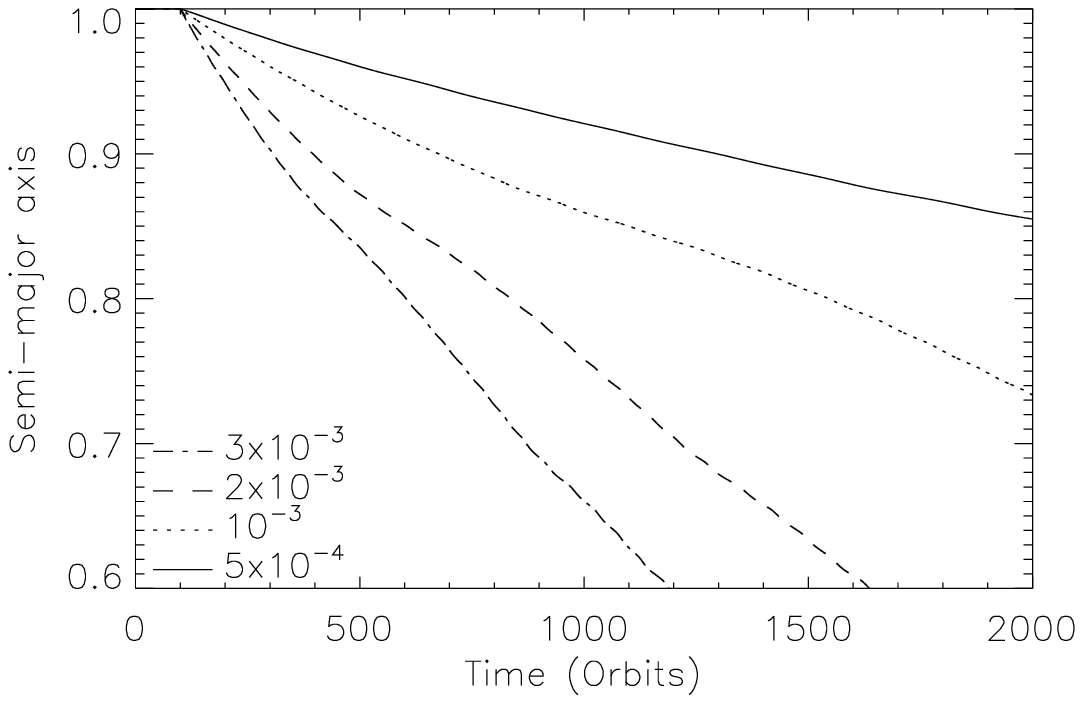}
\caption{Planetary migration in the $\nu=10^{-5}$ case for disc power laws of $\delta=0$ (top left), 0.5 (top right), 1.0 (bottom left) and 1.5 (bottom right). The planets were released to migrate after 100 orbits. The lines are marked by the value of \qdisc{} (see Equation~\ref{eq:qDiskDefine})}
\label{fig:MidViscMassCompare}
\end{figure*}

\begin{figure*}
\plottwo{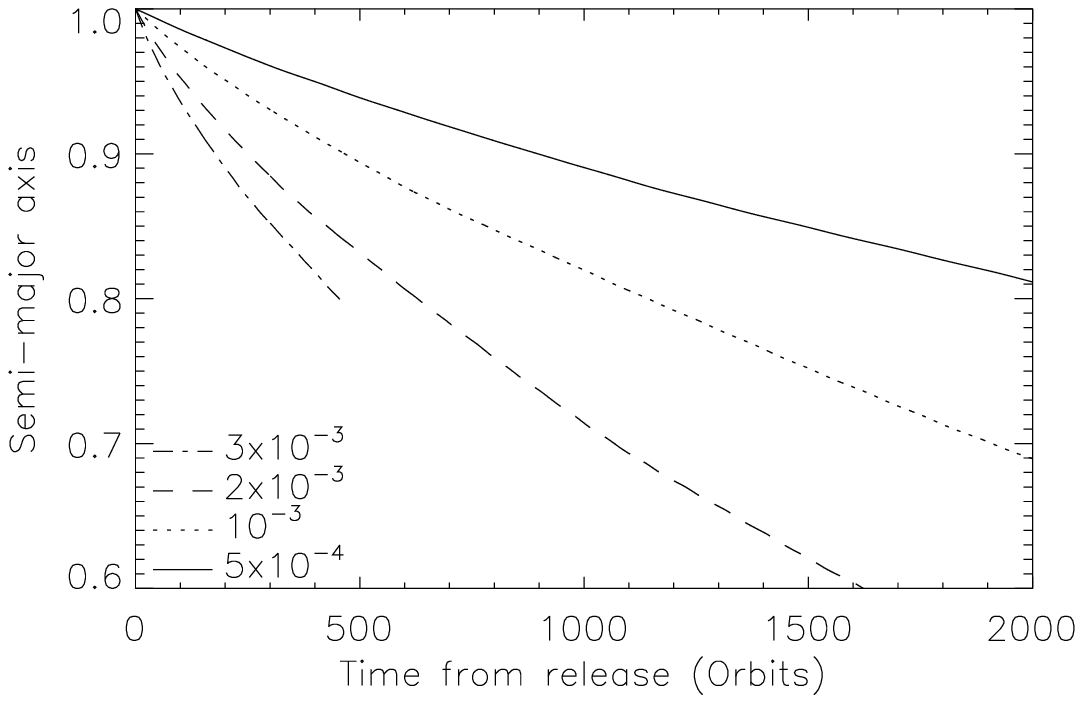}{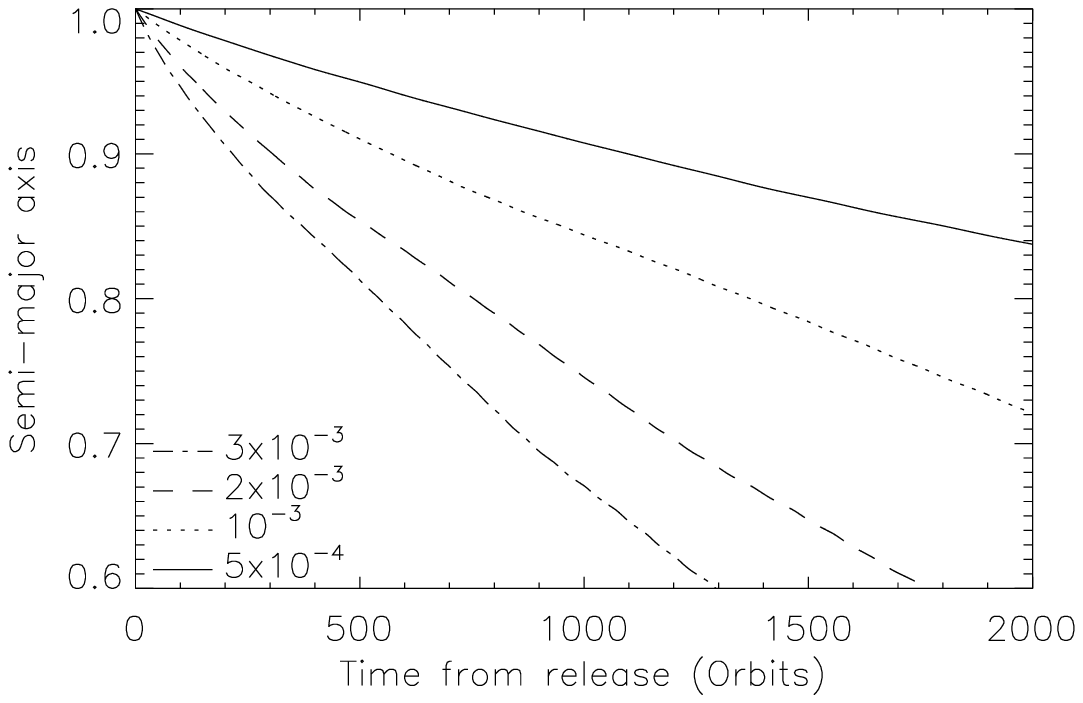} \\
\plottwo{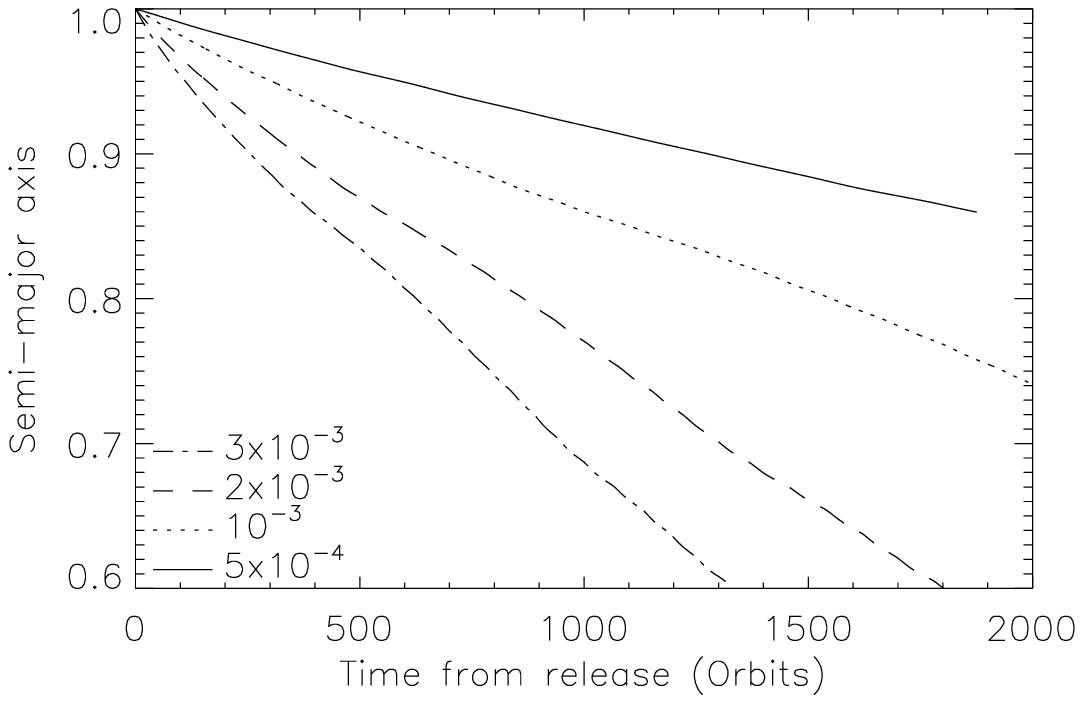}{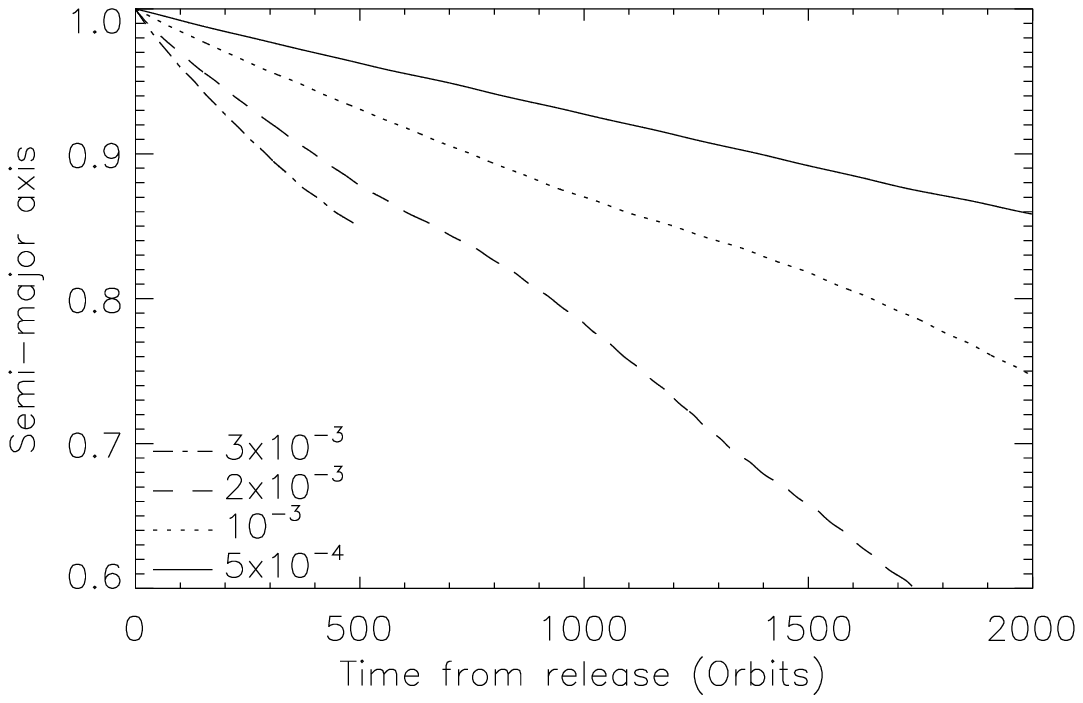}
\caption{Planetary migration in the $\nu=10^{-5}$ case for planets released after 1000 orbits. The disc power laws are $\delta=0$ (top left), 0.5 (top right), 1.0 (bottom left) and 1.5 (bottom right). The lines are marked by the value of \qdisc{} (see Equation~\ref{eq:qDiskDefine}). This plot should be compared to Figure~\ref{fig:MidViscMassCompare}}
\label{fig:MidViscMassCompareLongRelease}
\end{figure*}

\begin{figure}
\begin{center}
\plotone{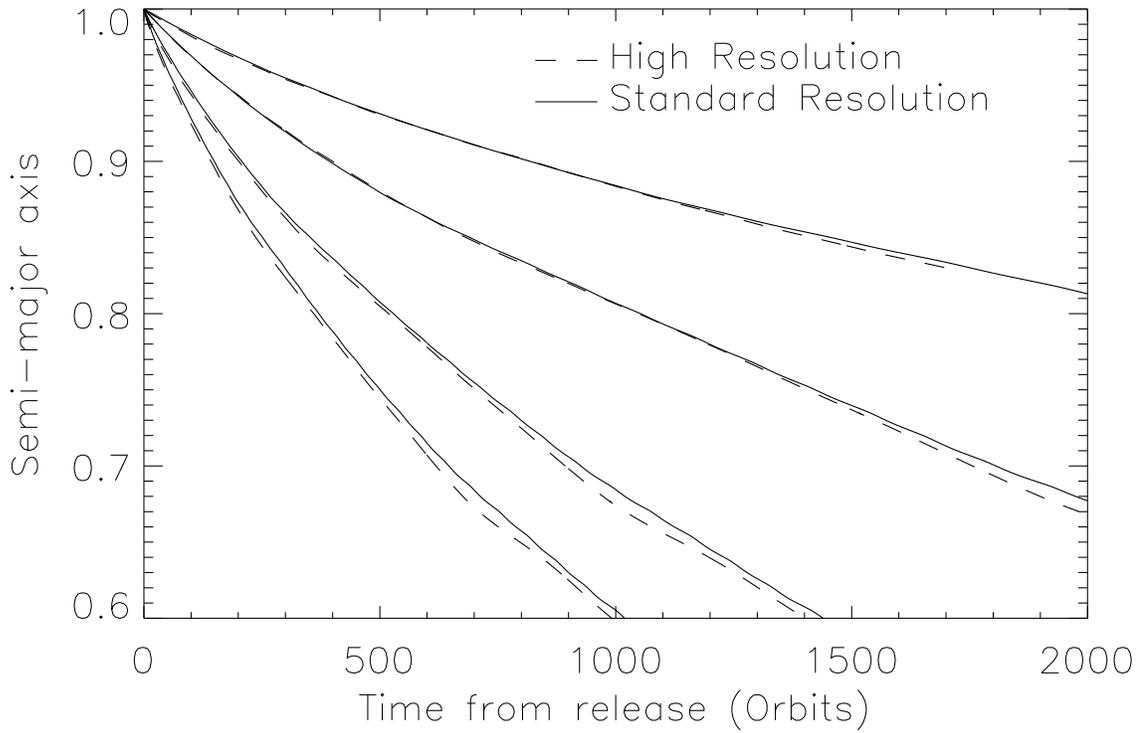}
\end{center}
\caption{Resolution test for a $\nu=10^{-5}$ disc with $\delta=0.5$, where the planet is released after 100 orbits. Four different \qdisc{} values are compared (refer to the top right panel of Figure~\ref{fig:MidViscMassCompare}) at two grid resolutions}
\label{fig:MidViscResolutionCompare}
\end{figure}

\begin{figure}
\begin{center}
\plotone{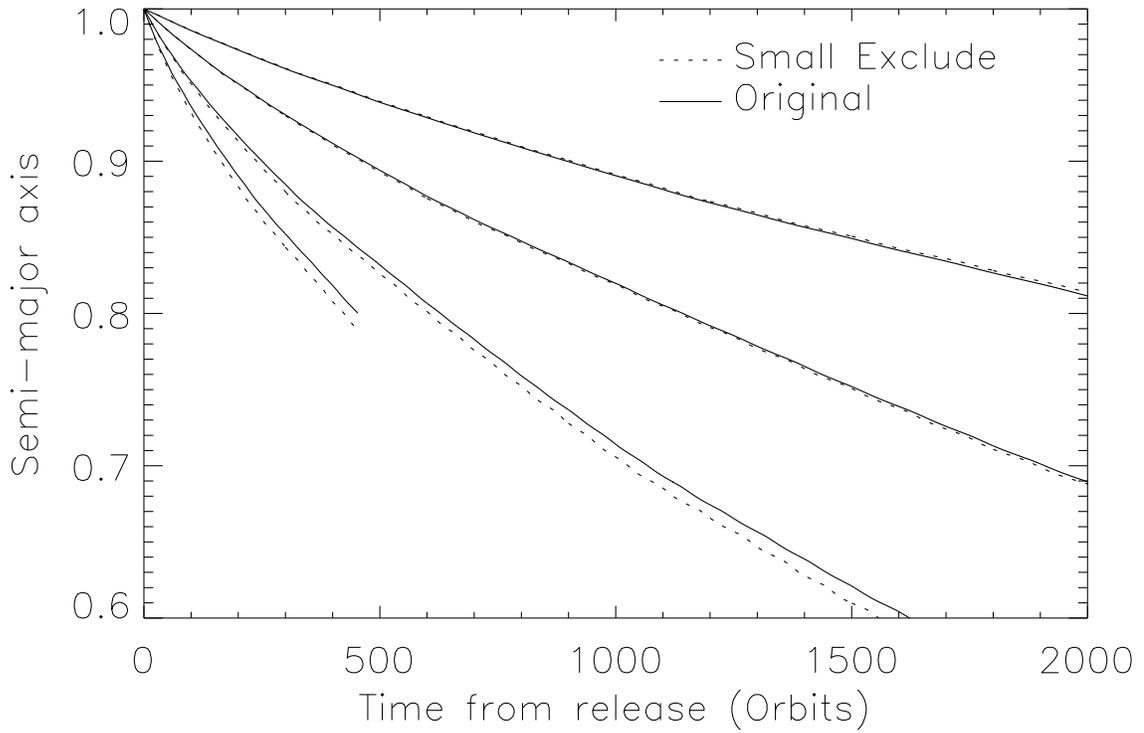}
\end{center}
\caption{Effect of reducing the exclusion radius to half the Hill radius when computing the migration torque on the planet. This figure is based on the $\delta=0$ (top left) panel of Figure~\ref{fig:MidViscMassCompare}, except that the planets were released after 1000 orbits. The standard four \qdisc{} migrations are plotted. The solid lines correspond to those in Figure~\ref{fig:MidViscMassCompare} (up to the difference in release time), while the dotted lines trace planets where the torque exclusion was only half the Hill radius}
\label{fig:MidViscRelease1kEvolve00exclusion}
\end{figure}

Figure~\ref{fig:MidViscRelease1kEvolve00exclusion} shows that this has negligible effect on our results.
This shows the migration of eight planets, embedded in $\nu=10^{-5}$, $\delta=0$ discs.
The four standard \qdisc{} values were considered, and the planets were held for 1000 orbits before being released to migrate.
The only difference between each pair of curves is whether the exclusion radius for the torque calculation was a full Hill radius, for the solid lines, or half the Hill radius, for the dotted lines.
The solid lines in figure~\ref{fig:MidViscRelease1kEvolve00exclusion} are equivalent to the $\delta=0$ (top left) panel of figure~\ref{fig:MidViscMassCompare}, up to the difference in release time.
In every case, the migration of each pair of planets is almost identical.

\begin{figure}
\plotone{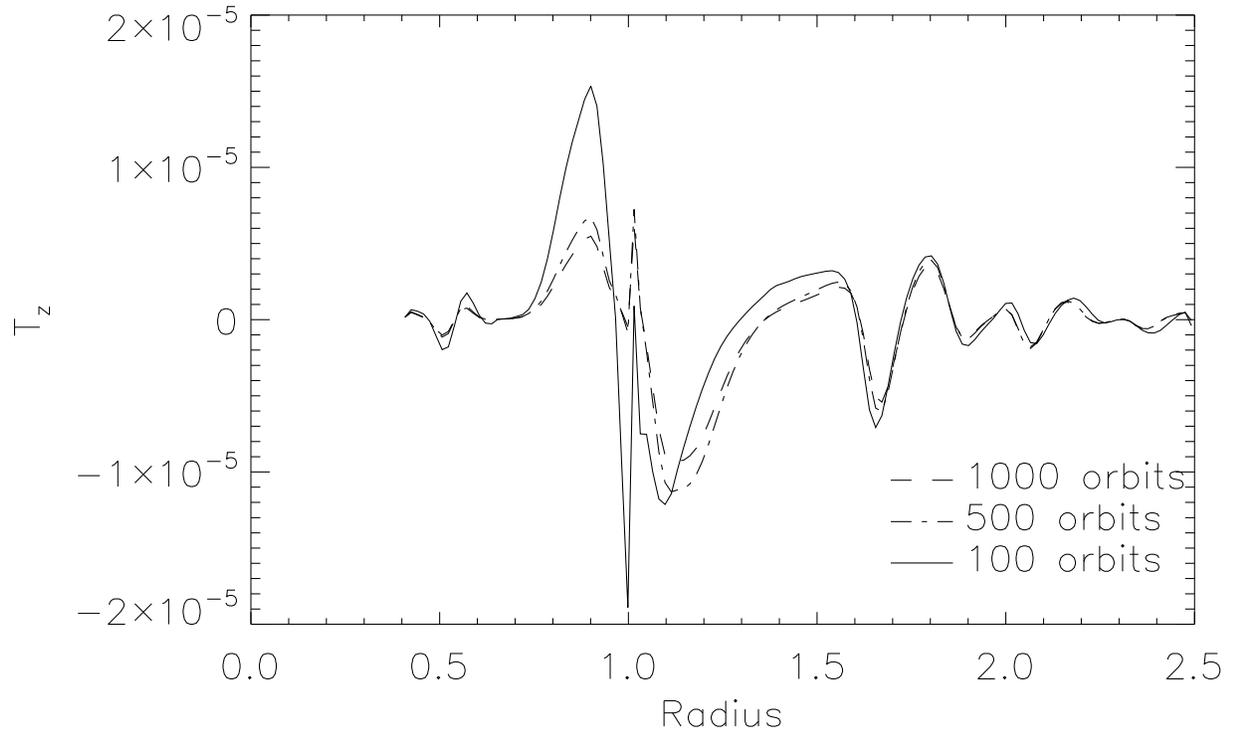}
\caption{Radial torque profile for a planet in a $\nu=10^{-5}$ disc as a function of time.
The disc initially had a flat ($\delta=0$) surface density profile, and the planet was on a fixed orbit for the entire time}
\label{fig:TorqueProfileFlatProfileEvolve}
\end{figure}

\begin{figure}
\begin{center}
\plotone{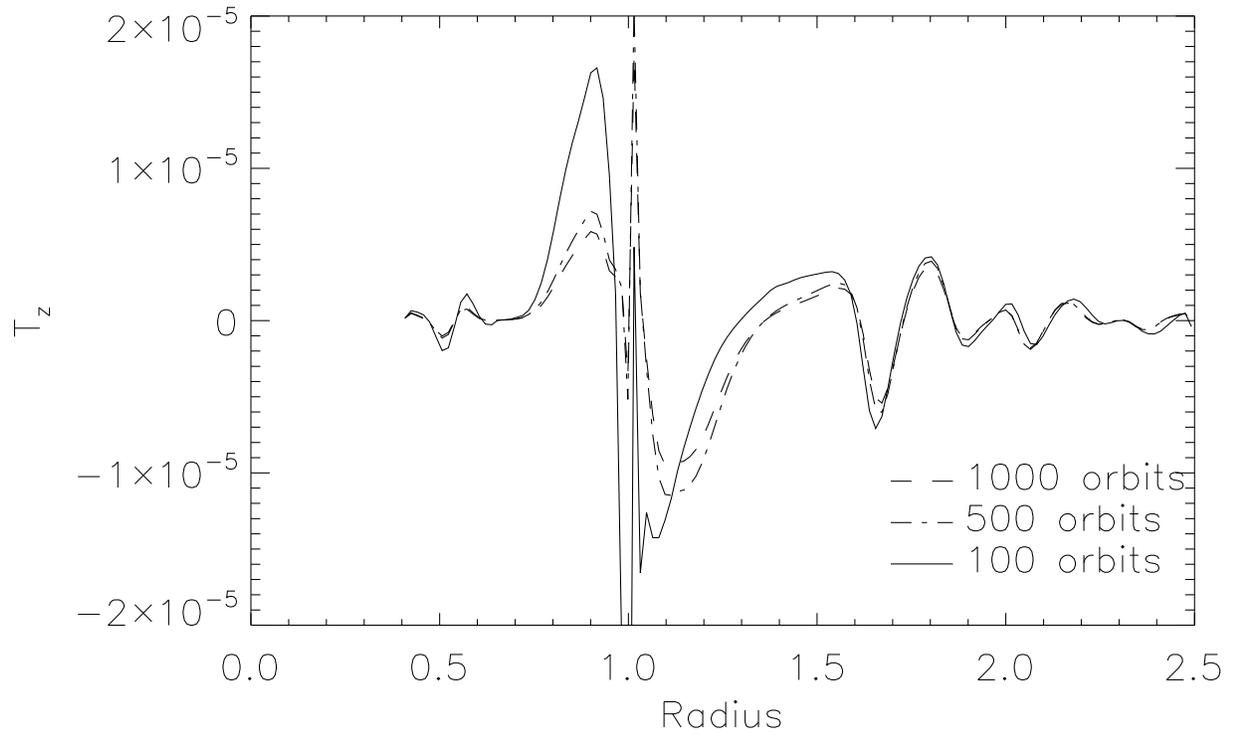}
\end{center}
\caption{Effect on the torque profiles of reducing exclusion radius to half the Hill radius. This figure should be compared to figure~\ref{fig:TorqueProfileFlatProfileEvolve}}
\label{fig:TorqueProfileFlatSmallExclude}
\end{figure}

In figure~\ref{fig:TorqueProfileFlatSmallExclude}, we show the effect of the exclusion radius reduction on the torque profiles.
This figure should be directly compared to figure~\ref{fig:TorqueProfileFlatProfileEvolve}.
We can see that the torque close to the planet is increased, particularly for the first curve, plotted after 100 orbits.
There is also a stronger peak close to the planet for all the curves.
Otherwise, the torque profiles are remarkably similar to figure~\ref{fig:TorqueProfileFlatProfileEvolve}.
This is as expected, given the results shown in figure~\ref{fig:MidViscRelease1kEvolve00exclusion}.


\acknowledgements

The author acknowledges support from NSF grants AST-0406799, AST-0098442, AST-0406823, and NASA grants ATP04-0000-0016 and NNG04GM12G (issued through the Origins of Solar Systems Program).
I would like thank Frederic Masset for use of the \textsc{Fargo} code.
I am also very grateful to Eric Blackman and Alice Quillen, for reading early drafts of this manuscript.
Some of the computations presented here used the resources of HPC2N, Ume\aa{}


\bibliography{general}
\bibliographystyle{astron}


\clearpage

\clearpage

\clearpage

\clearpage

\clearpage

\clearpage

\clearpage

\clearpage

\clearpage

\clearpage

\clearpage

\clearpage

\end{document}